**Electron Pairing and Nematicity in LaAlO$_3$/SrTiO$_3$ Nanostructures**

Aditi Nethwewala[1,2], Hyungwoo Lee[3], Jianan Li[1,2], Megan Briggeman[1,2], Yun-Yi Pai [1,2], Kitae Eom[3], Chang-Beom Eom[3], Patrick Irvin[1,2], Jeremy Levy[1,2*]

[1]Department of Physics and Astronomy, University of Pittsburgh, Pittsburgh, PA 15260.
[2]Pittsburgh Quantum Institute, Pittsburgh, PA, 15260.
[3]Department of Materials Science and Engineering, University of Wisconsin-Madison, Madison, WI 53706.

*Prof. Jeremy Levy, Department of Physics and Astronomy, University of Pittsburgh, Pittsburgh, PA 15260.

**Email address:** jlevy@pitt.edu

**Abstract**

Strongly correlated electronic systems exhibit a wealth of unconventional behavior stemming from strong electron-electron interactions. The LaAlO$_3$/SrTiO$_3$ (LAO/STO) heterostructure supports rich and varied low-temperature transport characteristics including low-density superconductivity, and electron pairing without superconductivity for which the microscopic origins is still not understood. LAO/STO also exhibits inexplicable signatures of electronic nematicity via nonlinear and anomalous Hall effects. Nanoscale control over the conductivity of the LAO/STO interface enables mesoscopic experiments that can probe these effects and address their microscopic origins. Here we report a direct correlation between electron pairing without superconductivity, anomalous Hall effect and electronic nematicity in quasi-1D ballistic nanoscale LAO/STO Hall crosses. The characteristic magnetic field at which the Hall coefficient changes directly coincides with the depairing of non-superconducting pairs showing a strong correlation between the two distinct phenomena. Angle-dependent Hall measurements further reveal an onset of electronic nematicity that again coincides with the electron pairing transition, unveiling a rotational symmetry breaking due to the transition from paired to unpaired phases at the interface. The results presented here highlights the influence of preformed electron pairs on the transport properties of LAO/STO and provide evidence of the elusive pairing "glue" that gives rise to electron pairing in SrTiO$_3$-based systems.

Many remarkable properties of electronic materials can be traced to the presence of strong electron-electron interactions and their coupling with other degrees of freedom. Unconventional superconductivity, various forms of magnetism, and electronic nematicity are some notable examples in this domain. Electronic nematicity is characterized by rotational symmetry breaking of an electronic fluid, resulting in strong anisotropic behavior which can be tunable with chemical potential or chemical doping, and by a magnetic field (1). Nematic phases have been found to exist in a wide range of electronic materials (2) extending from GaAs/AlGaAs heterostructures (3, 4) to high-temperature superconductors (5-7). Theoretical frameworks developed to help understand the origin of electronic nematicity face challenges because of the wide range of systems that exhibit this behavior (2). In strongly correlated systems such as high-temperature superconductors, electronic nematicity is often observed in the pseudogap regime (8). The precise connection between electronic nematicity and pseudogap behavior has empirical support but is not well established (5, 9-11).

SrTiO$_3$ (STO)-based heterostructures, and in particular formed with LaAlO$_3$ (LAO) (12) exhibit two-dimensional (2D) superconductivity without the need for chemical doping (13, 14). Prominent features include a characteristic dome shape of the superconducting critical temperature (15, 16), evidence for a pseudogap phase up to $T \approx 500$ mK (13), and (iii) evidence for electron pairing without superconductivity, seen in single-electron transistors (SETs) (17), and within quasi-1D ballistic nanowires (18-22). The characteristic magnetic field and temperature at which electrons unbind can be two orders of magnitude larger than the boundary for superconductivity. Hence, the paired-but-non-superconducting regime covers a significant region of parameter space, overlapping in temperature and magnetic field with a wide range of experiments performed on macroscopic devices (23). However, the influence of non-superconducting paired states on the transport properties has not been explored.

One class of macroscopic transport experiments involves anisotropic magnetoresistance (AMR) which has been explored by several groups (24-27). Here we summarize representative results by Joshua *et al.* (24). These experiments are performed in a Hall bar geometry in which an in-plane magnetic field $H^{||}$ is applied. Below a critical magnetic field, $H_c^{||}$ (24, 27) the anisotropy of the magnetoresistance is solely determined by the direction of magnetic field. However, above $H_c^{||}$, an additional component of anisotropy appears with the pinning of AMR along preferred directions (24). In the same parameter regime of carrier density and magnetic field, the onset of an anomalous Hall effect (AHE) is also observed (24). Hall measurements show a change in the slope of the Hall resistance, at the critical magnetic field, $H_c^{||}$. The magnitude of $H_c^{||}$ depends sensitively on the carrier density of the system (24). Both AMR and AHE have been linked to a Lifshitz transition (16) in which electrons from the $d_{xz}$ and $d_{yz}$ bands appear in addition to the lower energy $d_{xy}$ band. However, multiband theory cannot fully account for AMR and AHE (24, 28). In (24) the observed anisotropy and change in slope of the Hall response is ascribed to an emergence of magnetization at the interface due to breaking of Kondo singlets. However, the origin of the magnetic impurities leading to a Kondo phase has not been conclusively identified (18, 26-29).

The paired non-superconducting regime exists in the same region of carrier density and magnetic field as the reports of AMR and AHE (27). The pairing field ($B_p$) can vary between 1 T and reach values as high as 15 T (20). $B_p$ is also reported to increase with decreasing carrier density (17), consistent with the dependence of $H_c^{||}$ reported in (24). Hence, it is natural to ask if the preformed pairing phenomena and 2D AHE share an underlying physical basis.

Another factor influencing the electronic properties of STO-based heterostructures are ferroelastic domains (19, 30-33). Bulk undoped STO undergoes a ferroelastic transition from cubic to tetragonal crystal symmetry at $T \sim 105$ K, leading to the formation of ferroelastic domains (30) which are oriented along the X [100], Y [010] and Z [001] crystalline directions, and separated by nanometer-scale domain walls according to the domain tiling rules (34). Local probe measurements including scanning SQUID and scanning SET have revealed that transport at the LAO/STO interface is highly inhomogeneous, with current flowing preferentially along ferroelastic domain boundaries (30, 31, 33).

Transport measurements on mesoscopic devices created at the LAO/STO interface using conductive atomic force microscope (c-AFM) lithography (35) provide a powerful platform to explore the rich physics at the interface. Experiments by Pai *et al.* demonstrated a one-dimensional nature of electron pairing and superconductivity at the LAO/STO interface (36). The existence of Shubnikov-de Haas oscillations has also been linked to the magnetic depopulation of electron subbands in 1D systems (37) which can account for the widely observed mismatch between Hall carrier density measurements and those revealed by quantum oscillations (38). The existence of ballistic transport itself in quasi-1D geometries with a mean free path of $\sim 20$ µm (18), show signatures which are not obvious from macroscopic 2D measurements but possibly consistent with spatially resolved measurements. If ferroelastic domains, which usually decorate the LAO/STO interface, possess a network of 1D domain walls that percolate in 2D, it is indeed plausible that macroscopic transport behavior might be heavily influenced by the physics of these 1D channels. The high conductance of these edges, which have been demonstrated in numerous experiments, offers a way to connect the mesoscopic physics of artificial quasi-1D devices with the much larger set of experiments performed at macroscopic 2D interface.

Here we describe mesoscopic transport experiments which aim to probe the correlation between electron pairing, AHE, and electronic nematicity at the LAO/STO interface. These measurements are enabled by quasi-1D cross-shaped ballistic electron waveguides or "nanocrosses", created at the LAO/STO interface using c-AFM lithography (19). The nanocross devices serve as a building block to understand 1D electron physics at the LAO/STO interface. The multi-terminal nature of the nanocross allows four-terminal measurements to be performed simultaneously in both longitudinal and Hall configurations, allowing the two distinct physical phenomena to be directly compared (*Figure 1* (a)). Further, the role of anisotropy is investigated by changing the angle of the nanocross ($\varphi$) with respect to the [100] crystallographic direction (Figure 1 (a)). All nanocross devices are written at the same location on the sample unless mentioned otherwise.

The nanocross geometry, illustrated in *Figure 1* (a), is composed of two 1 µm-long crossed nanowire segments, with each of the four ends connected to two nanowire leads. Tunnel barriers (see Materials and Methods for details) of width $\sim 30$ nm isolate the nanocross from the two terminal leads, allowing the chemical potential to be uniformly tuned by either of the two available proximal side gates at a distance of $\sim 1$ µm, from the center of the nanocross, with voltage $V_{sg1}$ and $V_{sg2}$. The precise physical location of the side gates for LAO/STO nanostructures negligibly impacts the electronic structure within the conducting regions (18). A few volts of back gate voltage $V_{bg}$ is also used to provide coarse tuning of the chemical potential. Four-terminal longitudinal and Hall measurements are performed simultaneously as a function of the applied gate voltage $V_{sg}$ or chemical potential $\mu = \alpha V_{sg}$ where $\alpha$ is the measured lever arm (see Figure S1 for details), and as a function of an applied out-of-plane magnetic field, $\boldsymbol{B} = B\hat{z}$. All measurements are performed at or near the base temperature of the dilution refrigerator, $T \sim 50$ mK.

The zero-bias longitudinal conductance $G = dI/dV$ (*Figure 1*(b)) and transconductance $dG/d\mu$ (*Figure 1*(c)) is shown for Device A1 ($\varphi = 65°$) as a function of $\mu$, for magnetic fields ranging between $B = 0$ T and $B = 8$ T. Transport is quasi-ballistic with signatures of conductance quantization, similar to reports of straight ballistic electron waveguides (18). A conductance plateau near $G \approx 1.75 \ e^2/h$ appears at all magnetic fields. For magnetic fields larger than $B = 4$ T, the transition to this plateau broadens significantly, and a second plateau appears at $G \approx 0.90 \pm 0.05\ e^2/h$ at $B = 8$ T (*Figure 1* (b)). The corresponding line cuts for $dG/d\mu$ (*Figure 1* (c)) also shows a clear splitting of the *1.75 $e^2/h$* peak starting at $B = 4$ T. The fractional values of the conductance quantization steps is attributed to interference or scattering effects within the nanocross (19).

More insight into the electronic properties of the nanocross can be obtained from examining the transconductance, $dG/d\mu$, as a function of $\mu$ and $B$. Figure 2(c) shows the transconductance intensity map of Device A1, over the energy range $\mu = $ *1.90* meV to *2.60* meV and the full magnetic field range $-8$ T $< B < 8$ T. Analysis of the transconductance peak structure, which is overlaid, reveals a transition from a single peak to two peaks at a critical field $B_p$. Fitting of the split peaks above $B_p$ yields an estimate for $B_p \approx $ *3.9* T $\pm$ *0.4* T (Figure 2 (e)).

Next, we focus on the Hall measurements across the nanocross (Figure 2(d)). Hall measurements in quasi-1D systems have been widely explored in traditional semiconductors where a quenching of the Hall resistance is observed (39, 40). Hall measurements across quasi-1D nanocrosses is expected to highlight the microscopic origin of 2D Hall measurements reported at the LAO/STO interface. Figure 2(f) shows the field anti-symmetrized Hall resistance, $R_{xy}^{anti}$ averaged over the energy range $\mu = $ *1.95* meV to *2.55* meV for Device A1. Nonlinearities are observed in the Hall response as a function of magnetic field, $B$. $R_{xy}^{anti}$ vs $B$, shows similar trend at all side-gate potentials $V_{sg}$ (See Figure S2 for details). Fits to the intersection of the low-field and high-field asymptotes (Figure 2 (f)) yield a critical value at which the Hall coefficient changes: $B_H \approx $ *3.4* T $\pm$ *0.5* T.

Figure 3 shows Hall measurements performed as a function of the orientation of the nanocross with respect to the [100] crystallographic direction (denoted by angle $\varphi$ in Figure 3 (a)). $R_{xy}^{anti}$ is averaged over a small range of sidegate voltages for the magnetic field range $-8$ T $< B < 8$ T for three nominally identical Devices A1, A2, A3, where the nanocross is oriented at the same angle $\varphi = $ *65°*. Nearly identical S-shaped Hall nonlinearities are observed in all three devices (Figure 3 (b)), in which the Hall coefficient at low magnetic fields is higher than at high magnetic fields. The magnetic field at which the Hall resistance changes slope, labeled as $B_H$, coincides within measurement uncertainty for all three devices A1-A3.

The dependence of the Hall response on the nanocross angle $\varphi$ is summarized in Figure 3 (c). Four distinct angles between 0° and 90° are explored in Devices A1, B1, C, and D, oriented at $\varphi = $ *65°, 0°, 45°*, and $\varphi = $ *75°* respectively. Nonlinear Hall behavior is observed in all devices. However, the shape of the Hall response is found to depend strongly on nanocross orientation.

Table 1 summarizes the low-field Hall coefficient, $R_H^{low}$ and high-field Hall coefficient, $R_H^{high}$, with the separation between low and high being $B_H$, obtained from experiments with eight devices. Fitting procedures and additional Hall measurements for devices with $\varphi = $ *0°* and $\varphi = $ *65°* are shown in Figure S3 and Figure S4. Interestingly, all devices exhibit comparable values of Hall slope when $|B| < B_H$. The Hall transition field, $B_H$ is minimum for $\varphi = $ *45°*, $B_H \sim$ 1.8 T $\pm$ 0.2 T and maximum for $\varphi = $ *0°*, $B_H \sim$ 5.9 T $\pm$ 0.1 T within the range of error. The Hall slope for Device B2 could not be identified with the same degree of accuracy in the high field regime but $B_H \sim$ 5.2 T for magnetic field range, $-7$ T $< B < 7$ T (Figure S4(f)) similar to Device B1 also written at $\varphi \sim $ *0°*.

The pairing transition in electron waveguides is characterized by a new conductance plateau at $G = e^2/h$ between $G = 0$ and $G = 2e^2/h$, which takes place for $|B| = B_p$. For $|B| < B_p$, transport takes place via electron pairs; for $|B| > B_p$, the electron pair splits into spin-up and spin-down single-electron channels, with subband bottoms that split and show up as two distinct peaks in the transconductance map (Figure 2 (c)). The experimental results show that the field at which the Hall slope changes ($B_H$) coincides, within error, with electron pairing transition ($B_p = 3.9 \pm 0.4$ T). This relationship holds regardless of the value of the pairing field. Results on Device E Sample 2, summarized in Figure S5- S6 yield $B_p = 2.2 \pm 0.4$ T (Figure S6(e)) and $B_H = 2.4 \pm 0.6$T (Figure S6 (f)), which agree within the uncertainty of measurement.

The striking agreement between the pairing field and anomalous Hall response in LAO/STO heterostructures suggest an underlying physical mechanism that relates them. Previous explanations for anomalous Hall response were mainly restricted to single-particle descriptions involving multiple bands (16, 29, 41) or invoked magnetic interactions of unknown origins (24, 28). In (24) the anomalous Hall signature is described as a metamagnetic transition, an "emergence of magnetization" which occurs at a critical magnetic field in 2-15 Tesla range. Our experimental findings point to a specific origin of excess magnetization, one which is associated with the breaking of spin-singlet electron pairs. Above the pairing field, spin-singlet electron pairs unbind and spin-polarize, resulting in characteristic changes in the Hall response. This scenario was postulated in (27) but lacked a strong empirical basis. The results reported here provide direct evidence in support of this mechanism, associating the AHE with the pairing transition.

To understand the angular dependence of the Hall resistance across the nanocross (19), we need to consider the role of ferroelastic domains and their connection to the relevant d-orbital bands at the LAO/STO interface. Prior magnetotransport measurements on nanocross devices have revealed an inhomogeneous energy landscape which is nevertheless highly reproducible from one device to another (19) (similar to the three devices A1-A3 from Figure 3). The observed inhomogeneity (19) is attributed to a highly reproducible ferroelastic domain configuration artificially described at the LAO/STO interface by the nanocross devices. We extend this ferroelastic domain model provided in (19) for the four angles of the nanocross discussed earlier (Figure 4 (e-h)). For simplicity we only consider the lowest energy configuration. A clear variation is observed in the ferroelastic domain configuration, with $\varphi = 0°$ (Figure 4 (e)) and $\varphi = 45°$ (Figure 4 (h)) configurations forming the two extremes. While $\varphi = 0°$ signifies that the nanocross naturally coincides with the crystallographic axis, $\varphi = 45°$ nanocross is aligned parallel to the X-Y domain boundary, and $\varphi = 65°$ and $\varphi = 75°$ nanocrosses are intermediate between the two extreme configurations. The observed minima and maxima of the Hall transition field, $B_H$ also coincides with the two extreme ferroelastic domain configurations as summarized in Figure 4 and Table 1. As mentioned earlier, the pairing field in mesoscopic devices has been found to vary between 2 T and 15 T (20-22). The possible role of ferroelastic domains and domain boundaries in mediating electron pairing in LAO/STO-based nanostructures has also been previously suggested in (36). The results presented here give further empirical evidence linking the preformed electron pairs, AHE and ferroelastic domain structures in LAO/STO.

Angle dependence of Hall resistance shares an important aspect with AHE and AMR studies reported in literature (38). Above a critical magnetic field, they all exhibit a dramatic change in anisotropy or nonlinearity in the transport properties at the interface. Figure 5(a) shows the variation of $R_{xy}^{anti}$ as the magnetic field strength is increased from 1 T to 7 T for $0° \leq \varphi \leq 180°$. The graph assumes two axes of symmetry, rotational symmetry by 90° and mirror symmetry along 45°, and is interpolated between measured values. Figure 5(a) reveals the increase in variation of $R_{xy}^{anti}$ vs $\varphi$ with increasing magnitude of magnetic field. A non-monotonic trend is observed in $R_{xy}^{anti}$ versus $\varphi$ with higher harmonic in $\varphi$ as previously reported for in-plane AMR measurements (24-26, 42, 43).

To quantify the non-monotonic behavior of $R_{xy}^{anti}$, we define a measure of nematicity, $N(B)$ as the standard deviation of $R_{xy}^{anti}$ over the interval $0° < \varphi < 90°$. For an isotropic system, $N(B)$ is expected to be close to zero. However, Figure 5(b) shows that as the magnitude of magnetic field increases, there is an increase in the angular variation of the Hall response, which is quantified by the increasing magnitude of $N(B)$ indicating the onset of electronic nematicity.

As mentioned previously, AMR and emergence of nonlinearities in Hall resistance at a critical magnetic field $B_H$ has often been linked to a Lifshitz transition in which electrons from the $d_{xz}$ and $d_{yz}$ bands contribute to transport, in addition to the lower-energy $d_{xy}$ band (16, 29). The preformed pairs which are most likely composed of isotropic $d_{xy}$ carriers, dominate the low-field normal Hall response. The onset of anisotropic transport above the pairing field suggests that when electrons de-pair, they acquire $d_{xz}/d_{yz}$ characteristics which are known to be highly anisotropic. The pairing transition is thus consistent with the Lifshitz picture, but with a shift in electron orbitals coinciding with the pairing transition itself. This scenario also provides a plausible explanation for the consistent value of $R_H^{low}$ for all eight devices (see Table 1), since they are derived from the Hall response of the preformed $d_{xy}$ electron pairs. The value of $R_H^{low}$ presented here (see Table 1), also closely matches with the value of $R_H^{low}$ ($\approx$ 40 Ω/T) reported in (24) for 2D Hall bars with magnetic field applied out of plane.

In summary, simultaneous longitudinal and Hall measurements on quasi-1D ballistic nanocrosses sketched at the LAO/STO interface have revealed a direct correlation between the electron pairing transition and nonlinearities in the Hall response. Angle-dependent Hall measurements further show evidence of electronic nematicity whose onset also coincides with the pairing transition. A natural explanation is connecting the electron pairing transition to a shift between $d_{xy}$ electron pairs and $d_{xz}$ and $d_{yz}$ unpaired states, with the latter exhibiting a high degree of anisotropic behavior and nematicity. The correlation between electron pairing, AHE and electronic nematicity consolidates a wide range of seemingly disparate experimental findings reported in STO and construct a comprehensive understanding of the rich correlated nanoelectronics present in this system. The results presented in this work provide several new insights regarding this system; the elusive pairing "glue" in STO, the importance of the paired non-superconducting phase in the overall phase diagram of STO, and the need to look beyond single-particle descriptions.

## Materials and Methods

### Film fabrication

The LAO/STO heterostructures are epitaxially grown on TiO$_2$-terminated STO (001) substrates using pulser laser deposition. The thickness of LAO is precisely controlled by in-situ RHEED monitoring. To make a TiO$_2$-terminated substrate, as received STO substrates are etched with buffered HF for 1 min and annealed at *1000*°C for six hours. During the LAO growth, the substrate temperature is kept at *550*°C and oxygen partial pressure is *$10^{-3}$* mbar. LAO target is focused by KrF (*248* nm) excimer laser at a repetition rate of *3* Hz and a fluence of *1.8 $J/cm^2$*. After growth, the sample is slowly cooled down to room temperature under oxygen pressure of *1* atm.

### c-AFM lithography

Sixteen interface contacts, formed by milling 25 nm-deep trenches and subsequently depositing Ti/Au (4 nm/25 nm), surround a *25* μm x *25* μm "canvas" where devices are "sketched" with a voltage-biased c-AFM tip. Conducting paths are created by applying a positive bias $V_{tip} \sim 10$ V to the AFM tip, which locally protonates the LAO surface, thereby rendering the interface locally *n*-type conductive. An insulating state is locally restored by applying negative voltages to the tip ($V_{tip} \sim -3$ V). The nanocross is composed of two 1 μm-long crossed nanowire segments, created using a positive tip voltage $V_{tip} = 12$ V. Each arm of the nanocross has a tunnel barrier which is created by "erasing" with a negative tip voltage $V_{tip} = -4$ V over a distance $w_b = 30$ nm. The tunnel barriers decouple the nanocross from the two terminal leads, allowing the electron density of the nanocross to be tuned by a proximal side-gate, $V_{sg}$.

## Acknowledgments

We thank Bharat Jalan and Beena Kalisky for helpful discussions. CBE acknowledges support for this research through the Gordon and Betty Moore Foundation's EPiQS Initiative, Grant GBMF9065 and a Vannevar Bush Faculty Fellowship (ONR N00014-20-1-2844). Transport measurement at the University of Wisconsin–Madison was supported by the US Department of Energy (DOE), Office of Science, Office of Basic Energy Sciences (BES), under award number DE-FG02-06ER46327. JL acknowledges support from a Vannevar Bush Faculty Fellowship (ONR N00014-15-1-2847) and National Science Foundation (NSF PHY-1913034).

**References**

1. P. M. Chaikin, T. C. Lubensky, *Principles of Condensed Matter Physics* (Cambridge University Press, Cambridge, 1995), DOI: 10.1017/CBO9780511813467.
2. E. Fradkin, S. A. Kivelson, M. J. Lawler, J. P. Eisenstein, A. P. Mackenzie, Nematic Fermi Fluids in Condensed Matter Physics. *Annual Review of Condensed Matter Physics* **1**, 153-178 (2010).
3. M. P. Lilly, K. B. Cooper, J. P. Eisenstein, L. N. Pfeiffer, K. W. West, Evidence for an Anisotropic State of Two-Dimensional Electrons in High Landau Levels. *Physical Review Letters* **82**, 394-397 (1999).
4. R. R. Du *et al.*, Strongly anisotropic transport in higher two-dimensional Landau levels. *Solid State Communications* **109**, 389-394 (1999).
5. I. Božović, J. Levy, Pre-formed Cooper pairs in copper oxides and LaAlO3—SrTiO3 heterostructures. *Nature Physics* **16**, 712-717 (2020).
6. Y. Ando, K. Segawa, S. Komiya, A. N. Lavrov, Electrical resistivity Anisotropy from self-organized one dimensionality in high-temperature superconductors. *Physical Review Letters* **88** (2002).
7. V. Hinkov *et al.*, Electronic Liquid Crystal State in the High-Temperature Superconductor $YBa_2Cu_3O_{6.45}$. *Science* **319**, 597-600 (2008).
8. B. Keimer, S. A. Kivelson, M. R. Norman, S. Uchida, J. Zaanen, From quantum matter to high-temperature superconductivity in copper oxides. *Nature* **518**, 179-186 (2015).
9. M. Hashimoto, I. M. Vishik, R.-H. He, T. P. Devereaux, Z.-X. Shen, Energy gaps in high-transition-temperature cuprate superconductors. *Nature Physics* **10**, 483-495 (2014).
10. Q. Chen, J. Stajic, S. Tan, K. Levin, BCS–BEC crossover: From high temperature superconductors to ultracold superfluids. *Physics Reports* **412**, 1-88 (2005).
11. H. B. Yang *et al.*, Emergence of preformed Cooper pairs from the doped Mott insulating state in Bi2Sr2CaCu2O8+δ. *Nature* **456**, 77-80 (2008).
12. H. Y. Hwang, A. Ohtomo, N. Nakagawa, D. A. Muller, J. L. Grazul, High-mobility electrons in SrTiO3 heterostructures. *Physica E-Low-Dimensional Systems & Nanostructures* **22**, 712-716 (2004).
13. C. Richter *et al.*, Interface superconductor with gap behaviour like a high-temperature superconductor. *Nature* **502**, 528-531 (2013).
14. J. F. Schooley, W. R. Hosler, M. L. Cohen, Superconductivity in semiconducting $SrTiO_3$. *Physical Review Letters* **12**, 474-475 (1964).
15. A. D. Caviglia *et al.*, Electric field control of the $LaAlO_3/SrTiO_3$ interface ground state. *Nature* **456**, 624-627 (2008).
16. A. Joshua, S. Pecker, J. Ruhman, E. Altman, S. Ilani, A universal critical density underlying the physics of electrons at the $LaAlO_3/SrTiO_3$ interface. *Nature Communications* **3**, 1129 (2012).
17. G. Cheng *et al.*, Electron pairing without superconductivity. *Nature* **521**, 196 (2015).
18. A. Annadi *et al.*, Quantized Ballistic Transport of Electrons and Electron Pairs in $LaAlO_3/SrTiO_3$ Nanowires. *Nano Letters* **18**, 4473-4481 (2018).
19. A. Nethwewala *et al.*, Inhomogeneous energy landscape in LaAlO3/SrTiO3 nanostructures. *Nanoscale Horizons* **4**, 1194-1201 (2019).
20. M. Briggeman *et al.*, Pascal conductance series in ballistic one-dimensional $LaAlO_3/SrTiO_3$ channels. *Science* **367**, 769-772 (2020).

21. M. Briggeman *et al.*, One-dimensional Kronig–Penney superlattices at the LaAlO3/SrTiO3 interface. *Nature Physics* **17**, 1-6 (2021).
22. M. Briggeman *et al.*, Engineered spin-orbit interactions in LaAlO 3 /SrTiO 3 -based 1D serpentine electron waveguides. *Science Advances* **6**, eaba6337 (2020).
23. Y.-Y. Pai, A. Tylan-Tyler, P. Irvin, J. Levy, Physics of $SrTiO_3$-based heterostructures and nanostructures: a review. *Reports on Progress in Physics* **81**, 036503 (2018).
24. A. Joshua, J. Ruhman, S. Pecker, E. Altman, S. Ilani, Gate-tunable polarized phase of two-dimensional electrons at the $LaAlO_3$/$SrTiO_3$ interface. *Proceedings of the National Academy of Sciences* **110**, 9633 (2013).
25. A. Annadi *et al.*, Fourfold oscillation in anisotropic magnetoresistance and planar Hall effect at the $LaAlO_3$/$SrTiO_3$ heterointerfaces: Effect of carrier confinement and electric field on magnetic interactions. *Physical Review B* **87**, 201102 (2013).
26. M. Ben Shalom *et al.*, Anisotropic magnetotransport at the $SrTiO_3$/$LaAlO_3$ interface. *Physical Review B* **80**, 140403 (2009).
27. Y.-Y. Pai, A. Tylan-Tyler, P. Irvin, J. Levy, "$LaAlO_3$/$SrTiO_3$: a tale of two magnetisms" in Spintronics Handbook, Second Edition: Spin Transport and Magnetism. (CRC Press, 2019), vol. 2.
28. F. Gunkel *et al.*, Defect Control of Conventional and Anomalous Electron Transport at Complex Oxide Interfaces. *Physical Review X* **6**, 031035 (2016).
29. M. Ben Shalom, A. Ron, A. Palevski, Y. Dagan, Shubnikov-De Haas Oscillations in $SrTiO_3$/$LaAlO_3$ Interface. *Physical Review Letters* **105**, 206401 (2010).
30. M. Honig *et al.*, Local electrostatic imaging of striped domain order in $LaAlO_3$/$SrTiO_3$. *Nature Materials* **12**, 1112-1118 (2013).
31. B. Kalisky *et al.*, Locally enhanced conductivity due to the tetragonal domain structure in $LaAlO_3$/$SrTiO_3$ heterointerfaces. *Nature Materials* **12**, 1091-1095 (2013).
32. E. Persky *et al.*, Non-universal current flow near the metal-insulator transition in an oxide interface.
33. Y. Frenkel *et al.*, Anisotropic Transport at the $LaAlO_3$/$SrTiO_3$ Interface Explained by Microscopic Imaging of Channel-Flow over $SrTiO_3$ Domains. *ACS Applied Materials & Interfaces* **8**, 12514-12519 (2016).
34. F. W. Lytle, X-Ray Diffractometry of Low-Temperature Phase Transformations in Strontium Titanate. *Journal of Applied Physics* **35**, 2212-2215 (1964).
35. C. Cen, S. Thiel, J. Mannhart, J. Levy, Oxide nanoelectronics on demand. *Science* **323**, 1026-1030 (2009).
36. Y.-Y. Pai *et al.*, One-Dimensional Nature of Superconductivity at the $LaAlO_3$/$SrTiO_3$ Interface. *Physical Review Letters* **120**, 147001 (2018).
37. G. Cheng *et al.*, Shubnikov-de Haas-like Quantum Oscillations in Artificial One-Dimensional $LaAlO_3$/$SrTiO_3$ Electron Channels. *Physical Review Letters* **120**, 076801 (2018).
38. Y.-Y. Pai, A. Tylan-Tyler, P. Irvin, J. Levy, Physics of $SrTiO_3$-based heterostructures and nanostructures: a review. *Reports on Progress in Physics* **81**, 036503 (2018).
39. M. L. Roukes *et al.*, Quenching of the Hall Effect in a One-Dimensional Wire. *Physical Review Letters* **59**, 3011-3014 (1987).
40. G. Kirczenow, Hall effect and ballistic conduction in one-dimensional quantum wires. *Physical Review B* **38**, 10958-10961 (1988).

41. A. Fête *et al.*, Growth-induced electron mobility enhancement at the LaAlO$_3$/SrTiO$_3$ interface. *Applied Physics Letters* **106**, 051604 (2015).
42. H. J. H. Ma *et al.*, Giant crystalline anisotropic magnetoresistance in nonmagnetic perovskite oxide heterostructures. *Physical Review B* **95**, 155314 (2017).
43. K. Wolff, R. Eder, R. Schäfer, R. Schneider, D. Fuchs, Anisotropic electronic transport and Rashba effect of the two-dimensional electron system in (110) ${\mathrm{SrTiO}}_{3}$-based heterostructures. *Physical Review B* **98**, 125122 (2018).

**Figures and Tables**

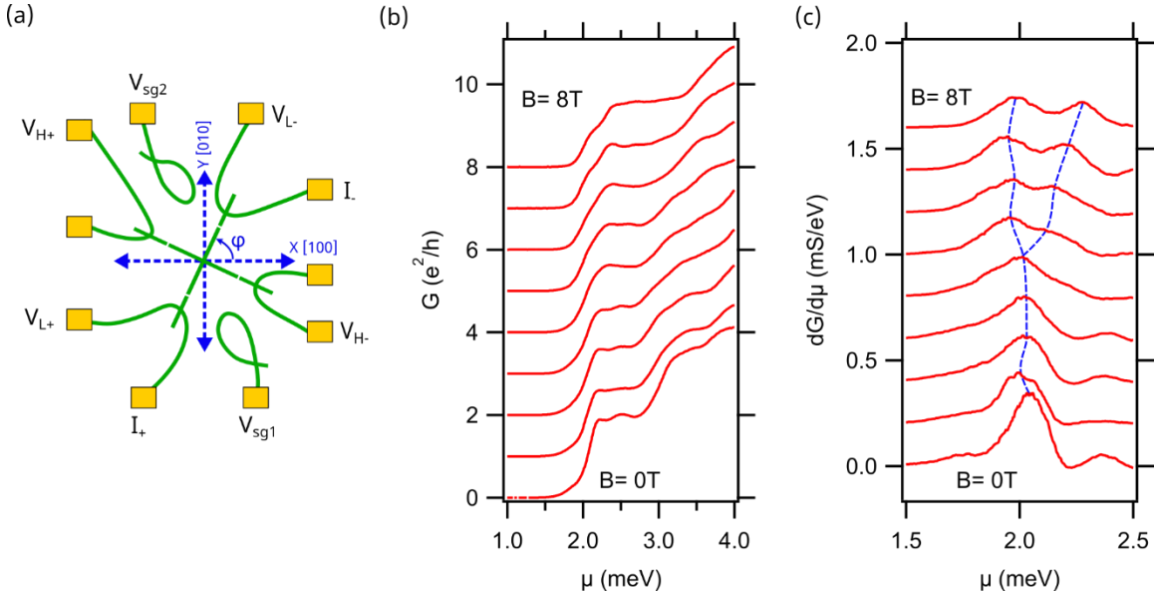

*Figure 1*: Nanocross geometry and longitudinal measurements across Device A1. (a) Schematic of longitudinal and Hall transport measurements across the nanocross. Longitudinal voltage probes ($V_{L\pm}$) enable four-terminal conductance to be measured while transverse voltage probes ($V_{H\pm}$) enable Hall measurements. Both longitudinal and Hall measurements are acquired simultaneously as a function of gate voltage and applied magnetic field, (b) Longitudinal conductance G versus chemical potential µ for magnetic fields ranging between B = 0 T and B = 8 T in steps of 1 T for Device A1 oriented at $\varphi = 65°$ with respect to [100] crystallographic direction. A conductance plateau near G ≈ 1.75 $e^2$/h appears at all magnetic fields. For magnetic fields larger than B = 4 T, the transition to this plateau broadens significantly, and a second plateau is clearly visible at G ≈ 0.90 ± 0.05 $e^2$/h at B = 8 T. Curves are offset by 1 $e^2$/h for clarity, (c) Transconductance dG/dµ versus µ for magnetic fields ranging between B = 0 T and B = 8 T in steps of 1 T for Device A1. dG/dµ versus µ reveals a transition between paired and unpaired state near B = 4 T as shown by the dashed blue lines. Curves are offset for clarity.

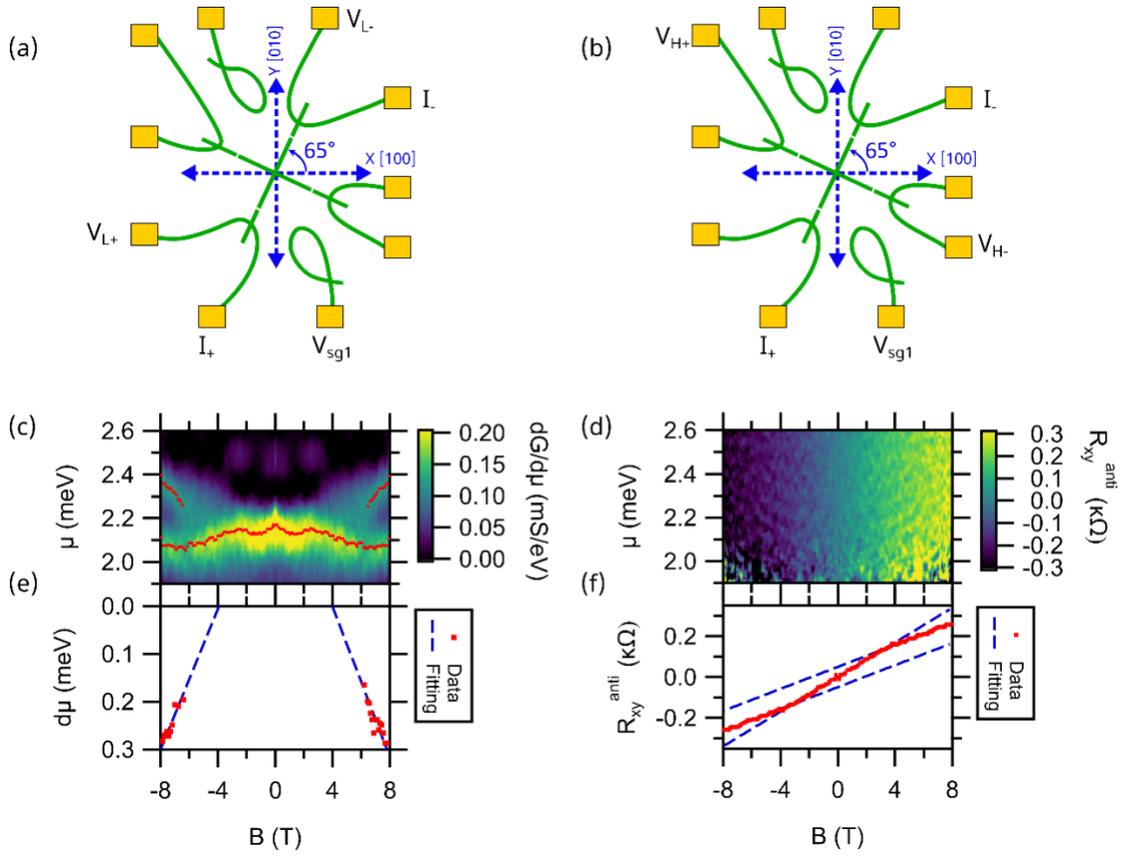

*Figure 2*: Comparison of transconductance $dG/d\mu$ and Hall measurements across Device A1. (a, b) Schematic showing the current and voltage lead configurations for longitudinal and Hall measurement across the nanocross. (c) Intensity plot of transconductance $dG/d\mu$ versus chemical potential μ and magnetic field $B$. Fits to peak of the transconductance versus magnetic field are overlaid. (d) Intensity plot of antisymmeterized Hall resistance $R_{xy}^{anti}$ versus $\mu$ and $B$. (e) Plot of energy difference between transconductance peaks versus magnetic field. Blue dashed line extrapolates to a value of $B_P = 3.9 \pm 0.4\,T$. (f) Average $R_{xy}^{anti}$ over the range $\mu = 1.95\,meV$ to $2.55\,meV$ reveals nonlinear behavior with asymptotes that cross at $B_H = 3.4 \pm 0.5\,T$.

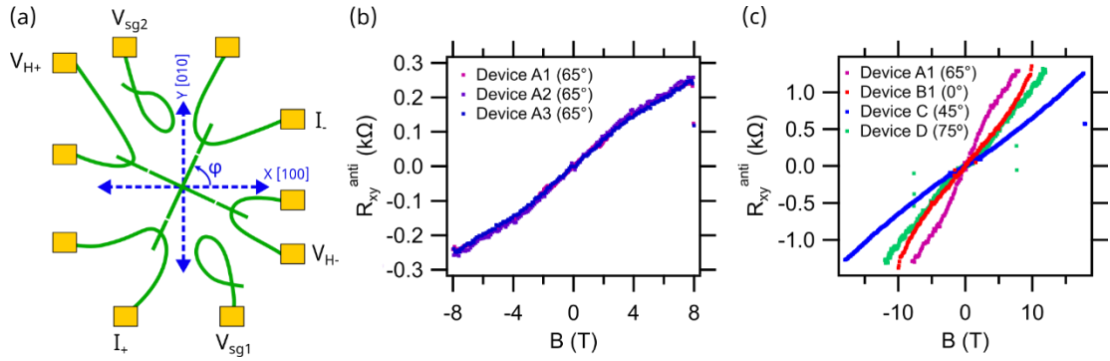

*Figure 3*: Angle dependence of anomalous Hall response. (a) Schematic showing the current and voltage leads for Hall measurement across the nanocross. Angle $\varphi$ denotes the relative position of the nanocross with respect to the crystallographic direction. (b) Hall measurements across Device A1, A2, A3 oriented at $\varphi = 65°$ with respect to the crystallographic direction. The Hall resistance overlaps within the uncertainty of measurement. (c) Variation of Hall resistance as a function of $\varphi$. Hall measurements across nanocross devices A1, B1, C and D oriented at $\varphi = 65°, 0°, 45°,$ and $75°$ respectively. Hall resistance for Device A1, B1 and D is amplified by 5x, 2.5x and 1.5x respectively for clarity.

*Table 1* : Slope of anomalous Hall response, transition field, $B_H$, and pairing field, $B_P$, across nanocross devices A1-E oriented at $\varphi = $ **0°, 45°, 65°**, and **75°**.

| Nanocross angle $\varphi$ | Sample | Device | $B_P$ (T) | $B_H$ (T) | $R_H^{low}$ (Ω/T) | $R_H^{high}$ (Ω/T) |
|---|---|---|---|---|---|---|
| **65°** | 1 | A1 | 3.9 ± 0.4 | 3.4 ± 0.5 | 43 | 26 |
| | | A2 | | 3.3 ± 0.4 | 42 | 24 |
| | | A3 | | 3.2 ± 0.3 | 42 | 24 |
| **0°** | | B1 | | 5.9 ± 0.1 | 43 | 74 |
| | | B2 | | 5.2 ± 2.9 | 44 | 37 |
| **45°** | | C | | 1.8 ± 0.2 | 49 | 70 |
| **75°** | | D | | 2.2 ± 0.1 | 49 | 76 |
| **45°** | 2 | E | 2.2 ± 0.4 | 2.4 ± 0.6 | 44 | 32 |

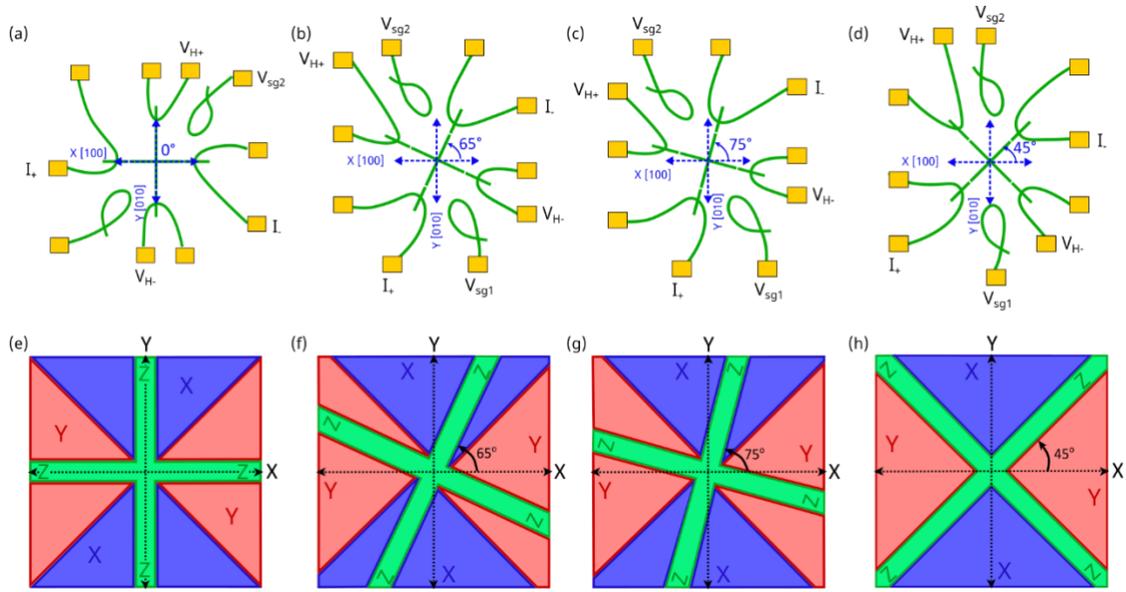

*Figure 4:* Ferroelastic domain model. (a-d) Schematic showing the current and voltage leads for Hall measurement across the nanocross devices oriented at $\varphi = 0°$, $65°$, $75°$ and $45°$ respectively, (e-h) The domain configuration of a symmetric nanocross in the lowest energy configuration for devices oriented at $\varphi = 0°$, $65°, 75°$ and $45°$ respectively. The Z-X, Z-Y and X-Y domain boundaries have been defined by darker shades along the edges of the nanocross for all cases.

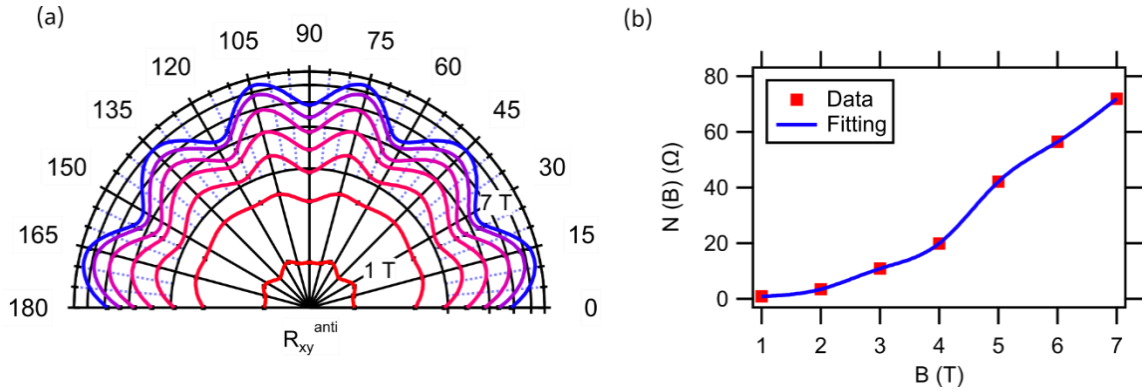

*Figure 5:* Angle dependence of Hall response and electron nematicity. (a) Spline fit showing variation in $R_{xy}^{anti}$ with increasing magnetic field strength, $1\,T \leq B \leq 7\,T$ for $0 \leq \varphi \leq 180°$, (b) Evolution of nematicity marker $N(B) = \langle \Delta R_{xy}^{anti\,2} \rangle^{1/2}$ as a function of magnitude of magnetic field $B$.

**Supplementary Information for**

**Electron Pairing and Nematicity in LaAlO$_3$/SrTiO$_3$ Nanostructures**

Aditi Nethwewala[1,2], Hyungwoo Lee[3], Jianan Li[1,2], Megan Briggeman[1,2], Yun-Yi Pai [1,2], Kitae Eom[3], Chang-Beom Eom[3], Patrick Irvin[1,2], Jeremy Levy[1,2*]

[1]Department of Physics and Astronomy, University of Pittsburgh, Pittsburgh, PA 15260, USA.
[2]Pittsburgh Quantum Institute, Pittsburgh, PA, 15260 USA.
[3]Department of Materials Science and Engineering, University of Wisconsin-Madison, Madison, WI 53706, USA.

*Prof. Jeremy Levy, Department of Physics and Astronomy, University of Pittsburgh.
 Email:  jlevy@pitt.edu

**This PDF file includes:**

    Supplementary text
    Figures S1 to S6
    SI References

## Supplementary Information Text

### Finite Bias Spectroscopy and "Lever arm" Ratio

The conversion factor that relates changes in gate voltage to changes in chemical potential is known as the "lever arm" ratio [1]. It can be calculated by analyzing the nonlinear current-voltage relation of a device as a function of the applied gate voltage. The lever arm ratio, $\alpha$ is defined by $\delta\mu = \alpha\, \delta V_{sg}$ where $\delta V_{sg}$ represents a change in the applied gate voltage and $\delta\mu$ denotes the resulting change in chemical potential. The horizontal red arrow in Figure S1 (b) marks the transition from one subband to another due to bias $\Delta V_{4T}$. The energy gain induced by $V_{4T}$ should be equal to the subband spacing marked by the vertical red arrow, $\alpha\Delta V_{sg}$ at zero bias, namely $e\Delta V_{4T} = \alpha\Delta V_{sg}$. Then $\alpha = eV_{4T}/\Delta V_{sg}$ can be precisely calculated [2]. For Device A1 shown in Figure S1(b), $\alpha = 5.0\ \mu eV/mV$.

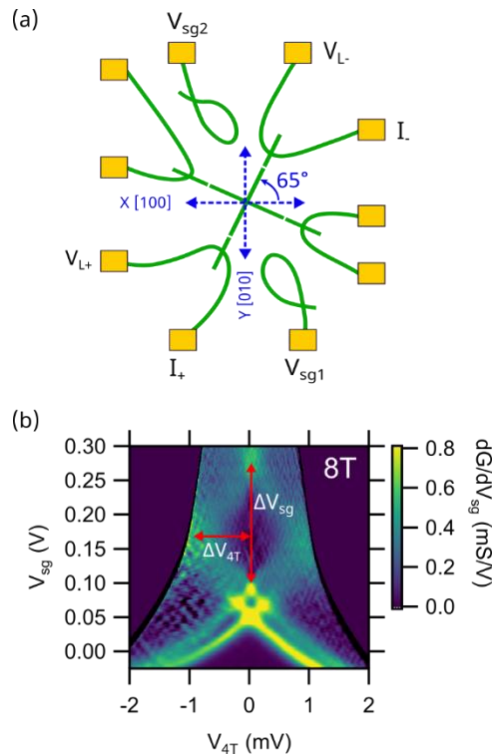

**Figure S1:** Finite Bias Spectroscopy on Device A1 (a) Schematic showing the device geometry, (b) The transconductance map, showing the diamond feature characteristic of ballistic transport. The red arrows denote the parameters used to calculate the lever arm.

**Variation of Hall resistance as a function of chemical potential for Device A1**

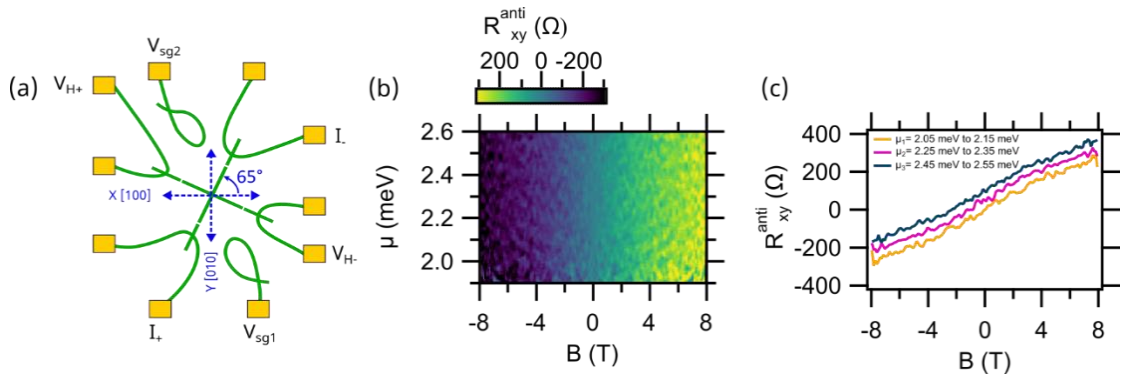

**Figure S2:** (a) Schematic showing the current and voltage leads for Hall measurement across the nanocross Device A1, (b) Field antisymmeterized Hall resistance, $R_{xy}^{anti}$, intensity map as a function of $\mu$ and $B$, (c) Line cuts of $R_{xy}^{anti}$ as a function of $B$ averaged between different ranges of chemical potential. The line cuts are shifted along the y-axis for clarity.

## Hall measurements across Devices A1, B1, C and D (Figure S3 (e-h))

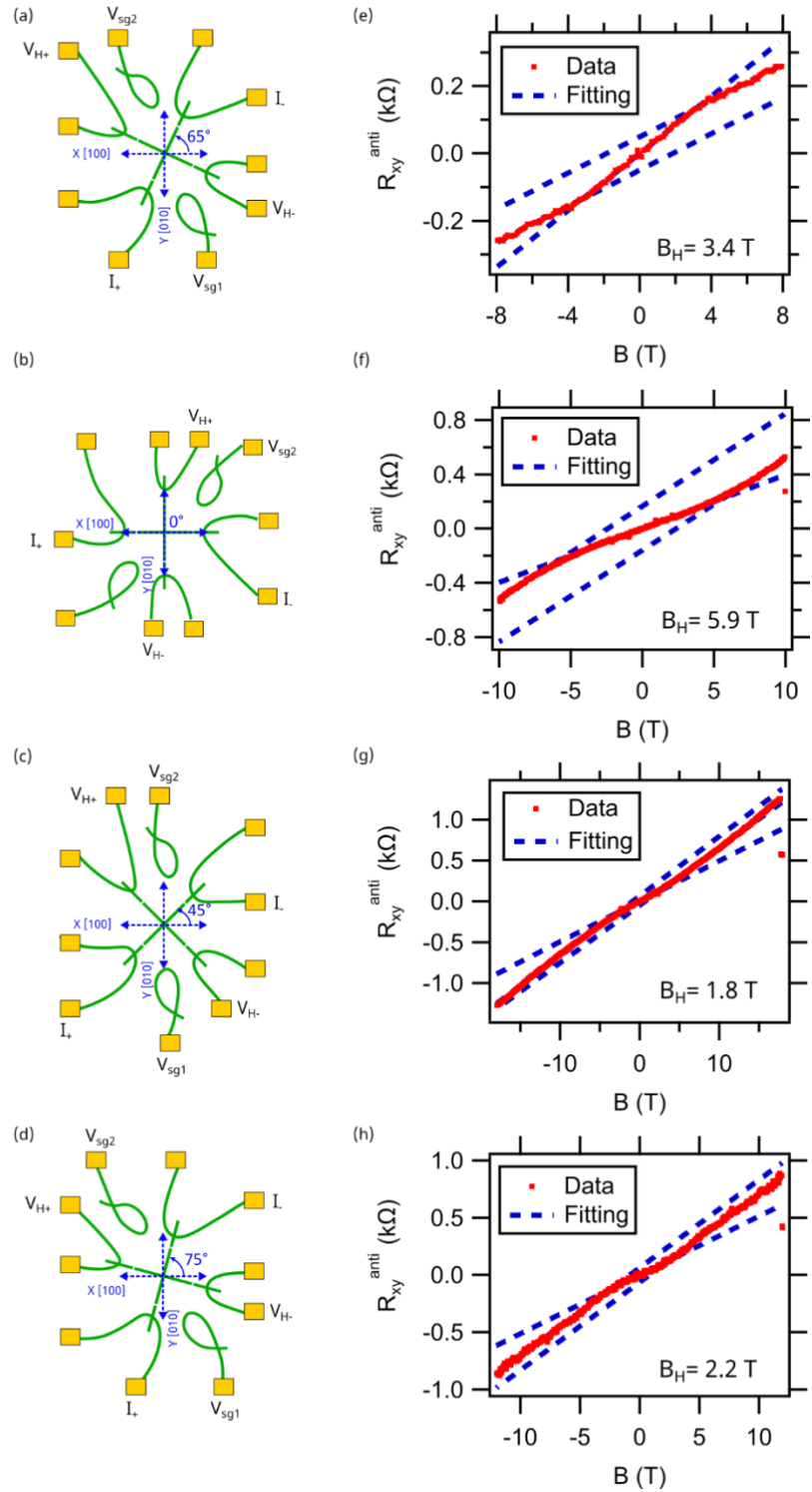

**Figure S3:** Hall measurements across nanocross devices A1, B1, C and D (a-d) Schematic showing the current and voltage leads for Hall measurement across nanocross devices A1, B1, C and D oriented at $\varphi = 65°, 0°, 45°$ and $75°$ respectively, (e-h) Hall resistance across nanocross devices A1, B1, C and D. Blue dashed lines show low-$B$ and high-$B$ asymptotes that cross at the Hall transition field $B_H$.

**Hall measurements across additional Devices A2, A3 and B2 (Figure S4 (d-f))**

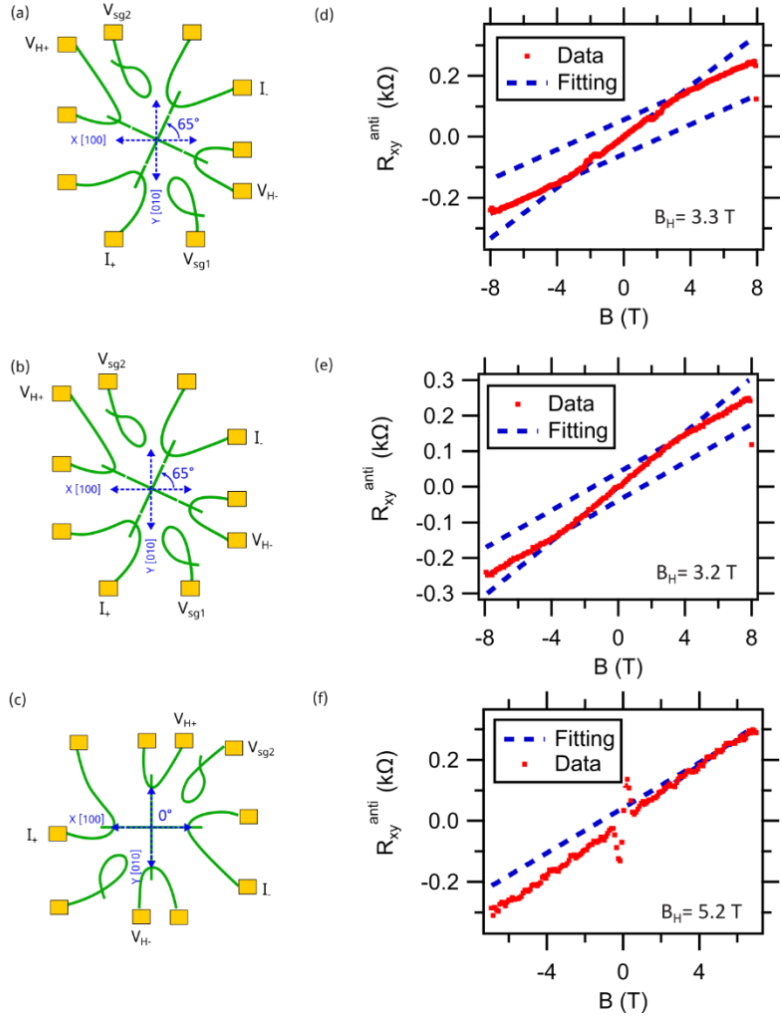

**Figure S4:** Hall measurements across nanocross devices A2, A3, B2. (a-c) Schematic showing the current and voltage leads for Hall measurement across nanocross devices A2, A3, and B2 oriented at $\varphi = 65°, 65°$, and $0°$, respectively, (d, e) Hall resistance across nanocross devices A2 and A3 oriented at $\varphi = 65°$, (f) Hall resistance across nanocross device B2 oriented at $\varphi = 0°$. Blue dashed lines show low-$B$ and high-$B$ asymptotes that cross at the Hall transition field $B_H$.

**Conductance and transconductance measurements on Device E**

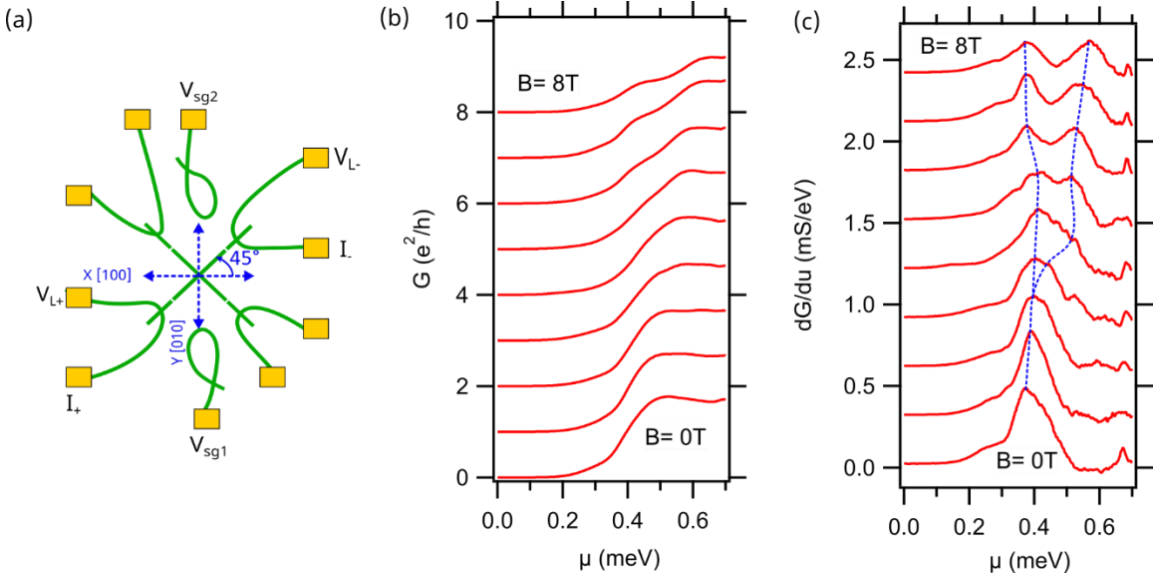

**Figure S5:** (a) Schematic of longitudinal and Hall transport measurements across the nanocross. Longitudinal voltage probes ($V_{L\pm}$) enable four-terminal conductance to be measured while transverse voltage probes ($V_{H\pm}$) enable Hall measurements. Both longitudinal and Hall measurements are acquired simultaneously as a function of gate voltage and applied magnetic field, (b) Longitudinal conductance $G$ versus chemical potential $\mu$ for magnetic fields ranging between $B = 0\,\text{T}$ and $B = 8\,\text{T}$ in steps of $1\,\text{T}$ on device E oriented at $\varphi = 45°$ on sample 2. A conductance plateau near $G \approx 1.70\, e^2/h$ appears at all magnetic fields. For magnetic fields larger than $B = 2\,\text{T}$, the transition to this plateau broadens significantly and a second plateau is clearly visible at $G \approx 0.90 \pm 0.05\, e^2/h$ at $B = 6\,\text{T}$. Curves are offset by $1\, e^2/h$ for clarity. (c) Transconductance $dG/d\mu$ versus $\mu$ for magnetic fields ranging between $B = 0\,\text{T}$ and $B = 8\,\text{T}$ in steps of $1\,\text{T}$. $dG/d\mu$ versus $\mu$ reveals a transition between paired and unpaired state near $B = 2\,\text{T}$ as shown by the dashed blue lines. Curves are offset for clarity.

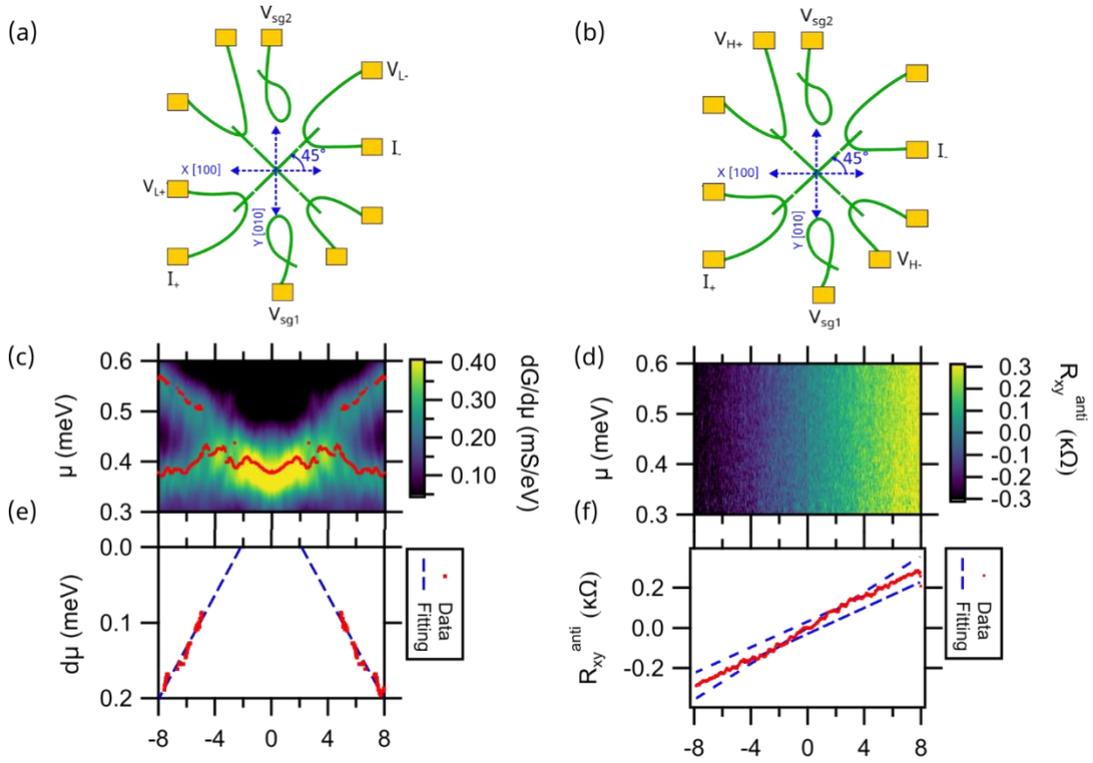

**Figure S6:** Comparison of transconductance $dG/d\mu$ and Hall measurements on Device E sample 2. (a, b) Schematic showing the current and voltage lead configurations for longitudinal and Hall measurement across nanocross device E. (c) Intensity plot of transconductance $dG/d\mu$ versus chemical potential $\mu$ and magnetic field $B$. Fits to peak of transconductance versus magnetic field are overlaid. (d) Intensity plot of Hall resistance $R_{xy}^{anti}$ versus $\mu$ and $B$. (e) Plot of energy difference between transconductance peaks versus magnetic field. Blue dashed line extrapolates to a value of $B_P = 2.2 \pm 0.4$ **T**. (f) Average Hall resistance over the range $\mu = $ **0.51 meV to 0.52 meV** reveals nonlinear behavior with asymptotes that cross at $B_H = 2.4 \pm 0.6$ **T**.

# References

[1] R.J. Warburton, B.T. Miller, C.S. Dürr, C. Bödefeld, K. Karrai, J.P. Kotthaus, G. Medeiros-Ribeiro, P.M. Petroff, S. Huant, Coulomb interactions in small charge-tunable quantum dots: A simple model, Physical Review B, 58 (1998) 16221-16231.
[2] A. Annadi, G. Cheng, H. Lee, J.-W. Lee, S. Lu, A. Tylan-Tyler, M. Briggeman, M. Tomczyk, M. Huang, D. Pekker, C.-B. Eom, P. Irvin, J. Levy, Quantized Ballistic Transport of Electrons and Electron Pairs in LaAlO$_3$/SrTiO$_3$ Nanowires, Nano Lett, 18 (2018) 4473-4481.